\documentclass[%
 amsmath,amssymb,
 aps,
]{revtex4-1}
\usepackage{graphicx}
\usepackage{dcolumn}
\usepackage{bm}

\begin{document}


\title{Some Aspects of Morris-Thorne Wormhole in Scalar Tensor Theory}

\author{Onur Gen\c{c}}
\email{onurgencnr@gmail.com}
\affiliation{%
 Department of Physics\\
Adnan Menderes University\\
Ayd\i n/TURKEY
}%



\begin{abstract}
In this study, we reach the equations of motion of Morris-Thorne wormhole geometry by means of the Einstein Field Equations and Klein-Gordon Equation of Scalar-Tensor theory by direct calculation. We discuss the anisotropic matter energy distribution. We determine a relation between the radial and the transverse pressures. Hence, we express the anisotropic energy momentum tensor in terms of one pressure class, by means of that relation. Besides that, we check the isotropic case and show that there is no traversable wormhole, in the zero redshift function situation, if the space is fulfilled with an isotropic energy momentum distribution. In addition, we represent the conditions in order that the  Null Energy Condition (NEC) is satisfied in the presence of a dilaton like field, in the zero tidal force case. We also propose a special type of shape function and express those conditions for it. We will be calling the wormhole corresponding to that shape function as the Yukawa Type. Furthermore, we determine the radial and the transverse pressures in the zero redshift function situation.
\end{abstract}

\pacs{Valid PACS appear here}
\maketitle


\section{\label{sec:level1}Introduction}

General Theory of Relativity (GTR) is inconsistent with the accelarated expansion of the universe phenomenon  \cite{GTRinconsistent} such that the present acceleration of the universe can only be explained with an enormously fine-tuned cosmological constant or with a completely ad hoc dark energy fluid as a matter source within the context of GTR \cite{CanOnlyBeExplainedinGTR}. This inconsistency leads to arise of the modified theories of gravitation which can be regarded as the extentions of the GTR. Among those extensions, possibly the simplest one that could accommodate the universe expansion acceleration are the Scalar Tensor Theories \cite{ExtensionOfGTRScalarTensorTheory} . In other words, Scalar Tensor theories are ones which are in agreement with the observation of the accelerated expansion of the universe. Besides that, Scalar Tensor theories can be regarded as the low energy limits of the quantum gravitational theories \cite{ScalarTensorTheoryLowEnergyLimit} and can be obtained as the low energy limit of string theories \cite{ScalarTensorTheoryLowEnergyString}.

Addition to basic reasons for studying in the scalar tensor theory it can be roughly declared that a scalar couples with Ricci scalar in a multiplicative way in those theories. In other words, this kind of theories are models of universes in which a scalar field and the space-time interacts. Very fundamental model of these Scalar Tensor theories is the Brans-Dicke (BD) model. The action functional of BD model is given as
\begin{eqnarray}
S\equiv\int d^{4}x (-g)^{1/2}\bigg\{ \psi R- \frac{\omega}{\psi}(\partial_{\sigma}\psi)^{2}-V(\psi) + \mathcal{L}_{m}   \bigg\}
\end{eqnarray}
where the scalar field potential is zero ($V(\psi)\equiv 0$) in the original theory \cite{BDreference}.  Here $\omega$ is only a parameter, for instance $\omega=-1$ which is a model-independent prediction of string theories \cite{ModelIndependentPredictionofStringTheories}, but not a general function of the field $\psi$ and the $\mathcal{L}_{m}$ is the lagrangian density corresponding to matter.

A generalised model of BD is expressed by the action functional
\begin{eqnarray} \label{Action Functional}
S&\equiv& \int d^{4}x(-g)^{1/2} \bigg\{f(\phi)\mathcal{L}_{m}(g_{\mu\nu},\xi)+\frac{1}{2\kappa}\bigg(\phi R- \frac{\omega(\phi)}{\phi}
(\partial_{\sigma}\phi)^{2}-V(\phi)\bigg)\bigg\} \nonumber\\
\end{eqnarray}
where $f(\phi)$ is an arbitrary function, $\kappa\equiv 8\pi G$, $G$ is the gravitational constant and $\xi$ corresponds to non gravitational fields \cite{ScalarFieldMatterLagrangian} (At this point and in the remaining part of this work we study in a unit system in which the vacuum speed of light is c=1). In this generalised model also a general function $\omega (\phi)$ of the scalar field $\phi$ shows up rather than the parameter $\omega$ as in the BD model.
In addition $f(\phi)=1$ in BD model.
If the action functional (\ref{Action Functional}) is extremized, the Einstein Field Equations and the Klein-Gordon Equation are obtained as below respectively \cite{ScalarFieldMatterLagrangian}
\begin{eqnarray} \label{Einstein Field Eqns.}
G_{\mu\nu}&\equiv &R_{\mu\nu}-\frac{1}{2}g_{\mu\nu}R=\kappa \frac{f(\phi)}{\phi} T_{\mu\nu}+\frac{1}{\phi}(\nabla_{\mu}\nabla_{\nu}-g_{\mu\nu}\Box)\phi+\frac {\omega(\phi)}{\phi^{2}}\bigg(-\frac{1}{2}g_{\mu\nu}(\partial_{\sigma}\phi)^{2}+(\partial_{\mu}\phi)(\partial_{\nu}\phi)\bigg)-g_{\mu\nu}\frac{V(\phi)}{2\phi}\nonumber\\
\end{eqnarray}
\begin{eqnarray} \label{Klein-Gordon Eqn.}
\frac{2\omega(\phi)+3}{\phi}\Box\phi &=& \kappa \bigg(\frac{f(\phi)}{\phi} T-2 f^{\prime}(\phi) \mathcal{L}_{m}\bigg)- \frac {2V(\phi)}{\phi}
-\frac{\omega^{\prime}(\phi)}{\phi}(\partial_{\sigma}\phi)^{2}+V^{\prime}(\phi)\nonumber\\
\end{eqnarray}
where prime denotes the partial derivative with respect to the scalar field $\phi$ (s.t. $W^{\prime}\equiv \partial_{\phi}W\equiv \partial W /\partial \phi$) and  $T$ denotes the trace of the energy momentum tensor. Equation (\ref{Einstein Field Eqns.}) is obtained from the variation of the action functional with respect to the metric tensor $g^{\mu\nu}$ and the equation (\ref{Klein-Gordon Eqn.}) is obtained from the variation with respect to the scalar field $\phi$. We use the above equation set (\ref{Einstein Field Eqns.}) and (\ref{Klein-Gordon Eqn.}) to obtain the equations of motion in our calculations.

Wormholes can be regarded as connections between two distinct points of space-time of either the same universe overlapped or two different universes. These two different sheets containing the upper and lower ends of the wormhole are called upper and lower sheets of the geometry respectively. Traversable wormholes in GTR were first obtained by Ellis and  Bronnikov \cite{EllisBronnikovWormhole} whereas the Einstein-Rosen bridge was regarded as nothing but a mathematical product since it is not traversable \cite{EinsteinRosenBridge}. Among those traversable ones, static, spherical symmetric Morris-Thorne wormhole is considered in our study. Basically, we mean by static here, the metric tensor components are independent of time. The geometry of Morris-Thorne wormhole is described by the metric
\begin{eqnarray} \label{TraversibleWormholeMetric}
ds^{2}&\equiv& -e^{2\Phi(r)}dt^{2}+\frac{dr^{2}}{1-b(r)/r}+r^{2}(d\theta^{2}+sin^{2}\theta d\zeta^{2})\nonumber\\
&\equiv&-e^{2\Phi(r)}dt^{2}+\frac{dr^{2}}{1-b(r)/r}+r^{2}d\Omega^{2}
\end{eqnarray}
where $b(r)$ determines the spatial shape of the wormhole, thus is called the shape function and $\Phi(r)$ determines the gravitational redshift, thus is called the redshift function \cite{Morris-Thorne}. The reader also should know that, here the radial coordinate $r$ is defined by the circumference $2\pi r$ of circles constructing up the wormhole. There exists a minimum value $r_{0}$ in the radial coordinate and this minimum value defines the throat of the wormhole, in particular the throat of the wormhole is the region where $r=r_{0}=r_{min.}$. The redshift function $\Phi$ must be finite everywhere satisfying the fact that there must not be horizons defined by the $e^{2\Phi}\rightarrow 0$ surfaces \cite{AnisotropicDistributionTensor} for a traversable wormhole. The absence of the horizons is obviously the required case, because we do not want regions from where we can not get any physical message for a geometry that is to be traversed through. Morever, $\Phi$ must satisfy $|\Phi|\ll1$ at the stations of travel in order that the gravitational redshift of the signals sent from the stations to the infinity is small where the stations are at far distances from the throat of the wormhole \cite{Morris-Thorne}. On the other hand, the conditions below must be hold by the shape function for a traversable wormhole
\begin{enumerate}
\item {$(b-b^{(1)}r)/b^{2}> 0$ (Flaring out condition)}

\item{$b(r=r_{0})= r = r_{0}$} (\underline{At the throat})

\item{$b^{(1)}(r_{0})<1$}

and

\item{$1-b(r)/r>0$}

\end{enumerate}
according to reference \cite{AnisotropicDistributionTensor} where $K^{(n)}\equiv \partial^{n}K/\partial r^{n}$.

In addition to fundamentals, we want to point out that, for further insight, one may also check the studies in which the wormhole geometry has been already discussed in scalar tensor theory. Among those, the stability of the static, spherically symmetric, traversable wormholes is studied in zero scalar potential class of scalar tensor theories in \cite{PastStudiesFirst} whereas the description of the wormhole by means of general static, spherically symmetric metric is included in  \cite{PastStudiesSecond}. In addition, it is shown, in the reference \cite{PastStudiesThird}  that zero Ricci scalar space times are solutions in scalar tensor theory with everywhere non zero $|g_{00}|$. Again for the zero Ricci scalar case, a connection between the gravitational lensing and the energy conditions is represented in \cite{PastStudiesFourth} asserting that the deflection of light is always positive if the energy conditions are satisfied.

Moreover, some other studies including the wormhole geometry in modifications of gravitation can be also checked for different perspectives. For instance, a class of spherically symmetric wormholes are constructed in quadratic $F(R)$ theory and it is expressed that, that construction forces a condition between the energy density, pressure and the tension, which can be regarded as the equation of state \cite{PastStudiesFifth}.  Furthermore, it is shown in reference \cite{PastStudiesSixth} that the existance of wormholes without the need of the exotic matter for an anisotropic fluid in few zones of the parameter space is possible in $f(R,T)$ gravity, whereas the dark wormholes are discussed in \cite{PastStudiesSeventh}, in $f(\tau)$ gravity, where $\tau$ is the torsion scalar.

\section{\label{sec:level1}{Equations of Motion}}

In the very first part of the calculations, we have determined the Ricci tensor components and the Ricci scalar by using the expressions for the Riemann tensor given in the reference \cite{Morris-Thorne} and checked the results benefiting from the general expressions for spherical symmetric metric
\begin{eqnarray} \label{GeneralizedSphericalSymmetricMetric}
ds^{2}&\equiv& -e^{2M(t,r)}dt^{2}+e^{2N(t,r)} dr^{2}+r^{2}d\Omega^{2}
\end{eqnarray}
represented in \cite{CarollLectureNotes} where $M(t,r)= \Phi(r)\;\;\&\;\;N(t,r)= H(r)=-\frac{1}{2}ln(1-b/r)$ in our case. We have also benefited from the Christoffel symbol expressions of \cite{CarollLectureNotes} in the next steps of the calculations.
 After that, we consider the presence of an anisotropic source in the universe. The energy momentum distribution threading the wormhole corresponding to that anisotropic case is expressed as $T_{\mu\nu}\equiv (\rho +p_{t})U_{\mu}U_{\nu}+ p_{t}g_{\mu\nu} +(p_{r}- p_{t} )  \chi_{\mu}\chi_{\nu}$ where $\chi^{\mu}\equiv (1-b(r)/r)^{1/2}\delta^{\mu}_{1}$ is the unit space-like vector in the radial direction, $\rho$ is the energy density, $p_{r}$ is the radial pressure measured in the direction of $\chi^{\mu}$ and  $p_{t}$ is the transverse pressure measured in the direction orthogonal to the direction of $\chi^{\mu}$ \cite{AnisotropicDistributionTensor}. The $\delta$ is to state the Kronecker's delta and $U^{\mu}$ is the 4-velocity of the source here. We will be in the co-moving frame (CMF), thus $U^{\mu}=(e^{-\Phi}, \vec{0})$ in our study. In addition, we choose $\mathcal{L}_{m}\equiv -\rho$ which is one possibility \cite{AnisotropicDistributionTensor}.

Via those conditions, from the Einsten field equations for $\mu=\nu=0,1,2$ cases we obtain
\begin{equation}\label{muequivnuequivzeroeqn}
\frac{b^{(1)}}{r^{2}}= \kappa\rho  \frac{f(\phi)}{\phi} +\frac {1}{2}e^{-2\Phi}\omega(\phi)\bigg({\frac{\dot{\phi}}{\phi}}\bigg)^{2}+  \frac{V(\phi)}{2\phi}
\end{equation}

\begin{eqnarray}\label{munuoneonesecond}
\frac {2\Phi^{(1)}e^{2\Phi}(1-b/r)}{r}-\frac{e^{2\Phi}b}{r^{3}}&=&\kappa e^{2\Phi} \frac{f(\phi)}{\phi}p_{r}+\frac{\ddot{\phi}}{\phi}+ \frac {\omega(\phi)}{2}{\bigg( \frac{\dot{\phi}}{\phi} \bigg)}^{2}-e^{2\Phi}\frac{V(\phi)}{2\phi}\nonumber\\
\end{eqnarray}

\begin{eqnarray} \label{munutwotwosecond}
&\;&e^{2\Phi}(1-b/r)\Phi ^{(2)}+e^{2\Phi}\bigg(\frac{(1-b/r)}{r}- \frac{(b^{(1)}r-b)}{2r^{2}}\bigg)\Phi^{(1)} +e^{2\Phi}(1-b/r)(\Phi^{(1)})^{2}-e^{2\Phi}\frac{(b^{(1)}r-b)}{2r^{3}}\nonumber\\
&\;&\qquad\qquad=\kappa e^{2\Phi} \frac{f(\phi)}{\phi}p_{t}+\frac{\ddot{\phi}}{\phi} + \frac {\omega(\phi)}{2}{\bigg( \frac{\dot{\phi}}{\phi} \bigg)}^{2}-e^{2\Phi}\frac{V(\phi)}{2\phi}\nonumber\\
\end{eqnarray}
equations respectively where dot denotes the total derivative with respect to the time for a field $\phi$ which is assumed to be space independent for which $\partial_{i}\phi=0$, $\forall i=1,2,3.$ In addition, the equation derived from the  $\mu=\nu=3$ case of the field equations is not represented due to the angular symmetry of our system which reduces the number of equations of motion by one. Moreover, one can easily notice here that $\rho$, $p_{r}$ $\&$ $p_{t}$ depend on both the time via the field $\phi$ and the functions of it and the radial co-ordinate via the remainings. In other words, the functions $\rho$, $p_{r}$ $\&$ $p_{t}$ in our study are not only the functions of the radial co-ordinate $r$, but also the functions of the time. Furthermore, the equations (\ref{muequivnuequivzeroeqn}), (\ref{munuoneonesecond}) $\&$ (\ref{munutwotwosecond}) respectively define those functions such that
\begin{equation}\label{rho}
\rho(r,t) = \frac{\phi}{ \kappa f(\phi)}\bigg[ \frac{b^{(1)}}{r^{2}}-\frac {1}{2}e^{-2\Phi}\omega(\phi)\bigg({\frac{\dot{\phi}}{\phi}}\bigg)^{2}-  \frac{V(\phi)}{2\phi}\bigg]
\end{equation}

\begin{eqnarray}\label{pr}
p_{r}(r,t)&=& \frac{\phi}{\kappa e^{2\Phi} f(\phi)}\bigg[\frac {2\Phi^{(1)}e^{2\Phi}(1-b/r)}{r}-\frac{e^{2\Phi}b}{r^{3}}-\frac{\ddot{\phi}}{\phi}- \frac {\omega(\phi)}{2}{\bigg( \frac{\dot{\phi}}{\phi} \bigg)}^{2}+e^{2\Phi}\frac{V(\phi)}{2\phi}\bigg]
\end{eqnarray}
\begin{eqnarray} \label{pt}
p_{t}(r,t)&=&\frac{\phi}{\kappa e^{2\Phi}f(\phi)} \bigg[  e^{2\Phi}(1-b/r)\Phi ^{(2)}+e^{2\Phi}\bigg(\frac{(1-b/r)}{r}- \frac{(b^{(1)}r-b)}{2r^{2}}\bigg)\Phi^{(1)} +e^{2\Phi}(1-b/r)(\Phi^{(1)})^{2}-e^{2\Phi}\frac{(b^{(1)}r-b)}{2r^{3}}\nonumber\\
&\;&\qquad\qquad\qquad\qquad -\frac{\ddot{\phi}}{\phi} - \frac {\omega(\phi)}{2}{\bigg( \frac{\dot{\phi}}{\phi} \bigg)}^{2}+e^{2\Phi}\frac{V(\phi)}{2\phi}\bigg]
\end{eqnarray}

which allows a dynamic energy momentum distribution throughout the scalar field, although our metric is static.

On the other hand, the Klein-Gordon equation asserts the
\begin{eqnarray}\label{phidotdoteqn.first}
\ddot{\phi}&=& e^{2\Phi} \frac{\kappa f(\phi)\rho }{2\omega(\phi)+3} \bigg(1 - \frac{ 2f^{\prime}(\phi)\phi}{f(\phi)} \bigg)-e^{2\Phi} \frac{\kappa f(\phi) }{2\omega(\phi)+3}(p_{r}+2p_{t})+ e^{2\Phi} \frac{\big(2V(\phi)-\phi V^{\prime}(\phi)\big)}{2\omega(\phi)+3}- \frac{\omega^{\prime}(\phi)}{2\omega(\phi)+3} {\dot{\phi}}^{2}
\end{eqnarray}
equation for the field. However, equation (\ref{muequivnuequivzeroeqn}) already defines the $\kappa f(\phi)\rho$, thus eliminating it we obtain the
\begin{eqnarray}\label{phidotdoteqn.second}
\ddot{\phi}&=&  \frac{e^{2\Phi} }{2\omega(\phi)+3} \bigg(1 - \frac{ 2f^{\prime}(\phi)\phi}{f(\phi)} \bigg)\bigg\{\frac{\phi b^{(1)}}{r^{2}}- \frac {1}{2}e^{-2\Phi}\phi \omega(\phi)\bigg({\frac{\dot{\phi}}{\phi}}\bigg)^{2}-\phi \frac{V(\phi)}{2\phi}\bigg\} -e^{2\Phi} \frac{\kappa f(\phi) }{2\omega(\phi)+3}(p_{r}+2p_{t})
\nonumber\\
&\;&\quad\quad\quad+ e^{2\Phi} \frac{\big(2V(\phi)-\phi V^{\prime}(\phi)\big)}{2\omega(\phi)+3}  - \frac{\omega^{\prime}(\phi)}{2\omega(\phi)+3} {\dot{\phi}}^{2}\nonumber\\
\end{eqnarray}
equation. Hence, inserting the above equation (\ref{phidotdoteqn.second}) in equations (\ref{munuoneonesecond}) and (\ref{munutwotwosecond}) we sum up with the
\begin{eqnarray}\label{munuoneonethird}
\frac {2\Phi^{(1)}e^{2\Phi}(1-b/r)}{r}-\frac{e^{2\Phi}b}{r^{3}}
&=&\kappa e^{2\Phi} \frac{f(\phi)}{\phi}p_{r}
+ \frac{e^{2\Phi} }{2\omega(\phi)+3} \bigg(1 - \frac{ 2f^{\prime}(\phi)\phi}{f(\phi)} \bigg)\bigg\{\frac{ b^{(1)}}{r^{2}}
- \frac {1}{2}e^{-2\Phi} \omega(\phi)\bigg({\frac{\dot{\phi}}{\phi}}\bigg)^{2}- \frac{V(\phi)}{2\phi}\bigg\}\nonumber\\
&\;&\quad-e^{2\Phi} \frac{\kappa f(\phi) }{\phi(2\omega(\phi)+3)}(p_{r}+2p_{t})
+ e^{2\Phi} \frac{\big(2V(\phi)-\phi V^{\prime}(\phi)\big)}{\phi(2\omega(\phi)+3)} - \frac{\omega^{\prime}(\phi)}{2\omega(\phi)+3} \frac{{\dot{\phi}}^{2}}{\phi}
\nonumber\\
&\;&\quad\quad\quad+ \frac {\omega(\phi)}{2}{\bigg( \frac{\dot{\phi}}{\phi} \bigg)}^{2}-e^{2\Phi}\frac{V(\phi)}{2\phi}
\end{eqnarray}
and
\begin{eqnarray} \label{munutwotwothird}
&\;&e^{2\Phi}(1-b/r)\Phi ^{(2)}+e^{2\Phi}\bigg(\frac{(1-b/r)}{r}- \frac{(b^{(1)}r-b)}{2r^{2}}\bigg)\Phi^{(1)} +e^{2\Phi}(1-b/r)(\Phi^{(1)})^{2}-e^{2\Phi}\frac{(b^{(1)}r-b)}{2r^{3}}\nonumber\\
&\;&\quad=\kappa e^{2\Phi} \frac{f(\phi)}{\phi}p_{t}+\frac{e^{2\Phi} }{2\omega(\phi)+3} \bigg(1 - \frac{ 2f^{\prime}(\phi)\phi}{f(\phi)} \bigg)\bigg\{\frac{ b^{(1)}}{r^{2}}
- \frac {1}{2}e^{-2\Phi} \omega(\phi)\bigg({\frac{\dot{\phi}}{\phi}}\bigg)^{2}-\frac{V(\phi)}{2\phi}\bigg\}\nonumber\\
&\;& \quad\quad\quad-e^{2\Phi} \frac{\kappa f(\phi) }{\phi(2\omega(\phi)+3)}(p_{r}+2p_{t})+ e^{2\Phi} \frac{\big(2V(\phi)-\phi V^{\prime}(\phi)\big)}{\phi(2\omega(\phi)+3)}- \frac{\omega^{\prime}(\phi)}{2\omega(\phi)+3} \frac{{\dot{\phi}}^{2}}{\phi}\nonumber\\
&\;&\quad\quad\quad\quad\quad
+ \frac {\omega(\phi)}{2}{\bigg( \frac{\dot{\phi}}{\phi} \bigg)}^{2}-e^{2\Phi}\frac{V(\phi)}{2\phi}
\end{eqnarray}
equations of motion. These above equations (\ref{munuoneonethird}) and (\ref{munutwotwothird}) are the general equations describing
the dynamics of our system.

\subsection{\label{sec:level2} Condition Relating the $p_{r}$ and $p_{t}$ and the New Form of the Energy Momentum Tensor}

We can also reach a general relation between the radial and the transverse pressures in terms of the field $\phi$, general function of the field $f(\phi)$, redshift function $\Phi$, the shape function $b$ and the radial coordinate $r$ as below
\begin{eqnarray}\label{PressureDifferenceGeneral}
p_{r}-p_{t}&=&\frac{\phi}{\kappa f(\phi)}\bigg\{\frac {2\Phi^{(1)}(1-b/r)}{r}-\frac{b}{r^{3}}- (1-b/r)\Phi ^{(2)}-\bigg(\frac{(1-b/r)}{r}- \frac{(b^{(1)}r-b)}{2r^{2}}\bigg)\Phi^{(1)} \nonumber\\
&\;&\quad\quad\quad\quad\quad-(1-b/r)(\Phi^{(1)})^{2}+\frac{(b^{(1)}r-b)}{2r^{3}}\bigg\}
\end{eqnarray}
by substracting the equation (\ref{munutwotwothird}) from the equation (\ref{munuoneonethird}).  One can also notice here that, the above equation (\ref{PressureDifferenceGeneral})
implies $p_{r}\rightarrow p_{t}$ as $r \rightarrow \infty$, in other words, the distribution in the universe is isotropic for large scales, for finite $\phi$, in accordance with the cosmological principle.

Therefore, we can express the anisotropic energy momentum tensor in terms of pressure measured in only one direction, rather than expressing it in terms of two pressures measured in the radial and the transverse directions distinctly as
\begin{eqnarray}\label{EnergyMomentumTensorOneComponentGeneral}
T_{\mu\nu}&=& (\rho +p_{t})U_{\mu}U_{\nu}+ p_{t}g_{\mu\nu}\nonumber\\
&\;&\quad+\frac{\phi}{\kappa f(\phi)}\bigg\{\frac {2\Phi^{(1)}(1-b/r)}{r}-\frac{b}{r^{3}}- (1-b/r)\Phi ^{(2)}-\bigg(\frac{(1-b/r)}{r}- \frac{(b^{(1)}r-b)}{2r^{2}}\bigg)\Phi^{(1)} \nonumber\\
&\;&\quad\quad\quad\quad\quad\quad\quad-(1-b/r)(\Phi^{(1)})^{2}+\frac{(b^{(1)}r-b)}{2r^{3}}\bigg\} \chi_{\mu}\chi_{\nu}\nonumber\\
\end{eqnarray}
by eliminating $(p_{r}-p_{t})$.
\section{\label{sec:level1}{Zero Tidal Force Case}}

The case of $\Phi=0$ corresponds to zero tidal force felt by stationary observers according to \cite{Morris-Thorne}.
Here, the tidal force effectively means the force between the top and the bottom points of a passenger passing through the wormhole.
Actually that force determines the limits for transferring via a traversable wormhole by a passenger.
Obviously, the tidal force must be necessarily small. In this section, we set $\Phi=0$, for which there is no such a discussion.
For this case, the equations of motion take the forms
\begin{eqnarray}\label{munuoneonefourth}
-\frac{b}{r^{3}}&=&\kappa \frac{f(\phi)}{\phi}p_{r}
+ \frac{1}{2\omega(\phi)+3} \bigg(1 - \frac{ 2f^{\prime}(\phi)\phi}{f(\phi)} \bigg)\bigg\{\frac{ b^{(1)}}{r^{2}}
- \frac {1}{2} \omega(\phi)\bigg({\frac{\dot{\phi}}{\phi}}\bigg)^{2}- \frac{V(\phi)}{2\phi}\bigg\}\nonumber\\
&\;&\quad-\frac{\kappa f(\phi) }{\phi(2\omega(\phi)+3)}(p_{r}+2p_{t})+ \frac{\big(2V(\phi)-\phi V^{\prime}(\phi)\big)}{\phi(2\omega(\phi)+3)}
- \frac{\omega^{\prime}(\phi)}{2\omega(\phi)+3} \frac{{\dot{\phi}}^{2}}{\phi}
+ \frac {\omega(\phi)}{2}{\bigg( \frac{\dot{\phi}}{\phi} \bigg)}^{2}-\frac{V(\phi)}{2\phi}
\end{eqnarray}
and
\begin{eqnarray} \label{munutwotwofourth}
-\frac{(b^{(1)}r-b)}{2r^{3}}&=&\kappa\frac{f(\phi)}{\phi}p_{t}+\frac{1}{2\omega(\phi)+3} \bigg(1 - \frac{ 2f^{\prime}(\phi)\phi}{f(\phi)} \bigg)\bigg\{\frac{ b^{(1)}}{r^{2}}
- \frac {1}{2} \omega(\phi)\bigg({\frac{\dot{\phi}}{\phi}}\bigg)^{2}-\frac{V(\phi)}{2\phi}\bigg\}\nonumber\\
&\;&\quad-\frac{\kappa f(\phi)
}{\phi(2\omega(\phi)+3)}(p_{r}+2p_{t})+ \frac{\big(2V(\phi)-\phi
V^{\prime}(\phi)\big)}{\phi(2\omega(\phi)+3)}-
\frac{\omega^{\prime}(\phi)}{2\omega(\phi)+3}
\frac{{\dot{\phi}}^{2}}{\phi} + \frac {\omega(\phi)}{2}{\bigg(
\frac{\dot{\phi}}{\phi} \bigg)}^{2}-\frac{V(\phi)}{2\phi}
\end{eqnarray}
respectively.

\subsection{\label{sec:level2}Non-Existance of a Traversable Wormhole with $\Phi=0$ in Isotropic Energy Momentum Distribution}

For an isotropic fluid, it can be asserted that $p_{r}= p_{t}= p$ \cite{IsotropicFluid}. With this condition our equations (\ref{munuoneonefourth}) and (\ref{munutwotwofourth}) are reduced to
\begin{eqnarray}\label{munuoneonefourthisotropic}
-\frac{b}{r^{3}}&=&\kappa \frac{f(\phi)}{\phi}p+ \frac{1}{2\omega(\phi)+3} \bigg(1 - \frac{ 2f^{\prime}(\phi)\phi}{f(\phi)} \bigg)\bigg\{\frac{ b^{(1)}}{r^{2}}
- \frac {1}{2} \omega(\phi)\bigg({\frac{\dot{\phi}}{\phi}}\bigg)^{2}- \frac{V(\phi)}{2\phi}\bigg\}\nonumber\\
&\;&\quad-\frac{3\kappa p f(\phi) }{\phi(2\omega(\phi)+3)}+ \frac{\big(2V(\phi)-\phi V^{\prime}(\phi)\big)}{\phi(2\omega(\phi)+3)}
- \frac{\omega^{\prime}(\phi)}{2\omega(\phi)+3} \frac{{\dot{\phi}}^{2}}{\phi}
+ \frac {\omega(\phi)}{2}{\bigg( \frac{\dot{\phi}}{\phi} \bigg)}^{2}-\frac{V(\phi)}{2\phi}
\end{eqnarray}
and
\begin{eqnarray} \label{munutwotwofourthisotropic}
-\frac{(b^{(1)}r-b)}{2r^{3}}&=&\kappa\frac{f(\phi)}{\phi}p+\frac{1}{2\omega(\phi)+3} \bigg(1 - \frac{ 2f^{\prime}(\phi)\phi}{f(\phi)} \bigg)\bigg\{\frac{ b^{(1)}}{r^{2}}
- \frac {1}{2} \omega(\phi)\bigg({\frac{\dot{\phi}}{\phi}}\bigg)^{2}-\frac{V(\phi)}{2\phi}\bigg\}\nonumber\\
&\;&\quad\quad-\frac{3\kappa p f(\phi) }{\phi(2\omega(\phi)+3)}+ \frac{\big(2V(\phi)-\phi V^{\prime}(\phi)\big)}{\phi(2\omega(\phi)+3)}- \frac{\omega^{\prime}(\phi)}{2\omega(\phi)+3} \frac{{\dot{\phi}}^{2}}{\phi}
+ \frac {\omega(\phi)}{2}{\bigg( \frac{\dot{\phi}}{\phi} \bigg)}^{2}-\frac{V(\phi)}{2\phi}
\end{eqnarray}
respectively, as a special case of the anisotropic situation. If we analyze the above equations (\ref{munuoneonefourthisotropic}) and (\ref{munutwotwofourthisotropic}) at this point, we see that they require
\begin{equation}\label{bEquation}
-\frac{b}{r^{3}}=-\frac{(b^{(1)}r-b)}{2r^{3}}
\end{equation}
because their right hand sides (RHSs) are exactly the same. The above differential equation (\ref{bEquation}) can be easily solved and the shape function is obtained as
\begin{equation}
b(r)=c_{I}r^{3}
\end{equation}
where $c_{I}$ is the integral constant. However, that shape function does not satisfy the conditions of traversable wormhole shape function. For example,
$b(r=r_{0})=c_{I}{r_{0}}^{3}\neq r_{0}$ at the throat of the wormhole except that $c_{I}=1/{r_{0}}^{2}$, but the flaring out condition is not satisfied for that case.
Therefore, no matter the integration constant is (even if it equals to $1/{r_{0}}^{2}$), $b(r)=c_{I}{r}^{3}$ is not a shape function.
 Hence, we can assert that there is no traversable wormhole,
 for which the tidal force is zero, in the presence of isotropic fluid in the universe in the model described by the action functional (\ref{Action Functional}).
 Actually for the dust source for which $T_{\mu\nu}\equiv \rho U_{\mu}U_{\nu}$, which is already a special zero pressure case of the isotropic distribution, the same situation holds, because of the fact that the RHSs of the equations of that case corresponding to equations (\ref{munuoneonefourthisotropic}) and (\ref{munutwotwofourthisotropic}) expressed here are exactly the same again.

\subsection{\label{sec:level2}Null Energy Condition}

The null energy condition claims that, $T_{\alpha \beta}N^{\alpha}N^{\beta}\geq 0$, $\forall N^{\gamma}$ such that $N^{\alpha}N_{\alpha}=0$ \cite{NEC}.
On the other hand, it is argued in the Appendix A of \cite{ScalarFieldMatterLagrangian}, performing a transformation, the action functional (\ref{Action Functional})
is the same with the string tree level effective action with $f(\phi)=\phi$ which corresponds to Damour and Polyakov's dilaton,
although there exists a convergence to general relativity problem. In this dilaton like case, equation (\ref{PressureDifferenceGeneral}) and equation (\ref{EnergyMomentumTensorOneComponentGeneral}) reads
\begin{equation}\label{PressureDifferenceSpecialCase}
p_{r}-p_{t}=\frac{-3b+b^{(1)}r}{2\kappa r^{3}}
\end{equation}
and
\begin{equation}\label{EnergyMomentumTensorOneComponentDilaton}
T_{\mu\nu}= (\rho +p_{t})U_{\mu}U_{\nu}+ p_{t}g_{\mu\nu} +\frac{(-3b+b^{(1)}r)}{2r^{3}}  \chi_{\mu}\chi_{\nu}
\end{equation}
respectively, for $\Phi=0$. If we propose a null vector $N^{\gamma}=(3^{1/2}e^{-\Phi}, (1-b/r)^{1/2}, r^{-1}, r^{-1}(sin\theta)^{-1})$ in general and
$N^{\gamma}=(3^{1/2}, (1-b/r)^{1/2}, r^{-1}, r^{-1}(sin\theta)^{-1})$ in this section, according to our wormhole metric, the NEC, described briefly above, imposes
\begin{equation} \label{NECconditionOne}
\rho+p_{t}+\frac{-3b+b^{(1)}r}{6\kappa r^{3}} \geq 0
\end{equation}
and
\begin{equation} \label{NECconditionTwo}
\rho+p_{r}-\frac{(-3b+b^{(1)}r)}{3\kappa r^{3}} \geq 0
\end{equation}
by means of equation (\ref{PressureDifferenceSpecialCase}).

\subsection{\label{sec:level2}{Conditions for Yukawa Type Wormhole}}

In this subsection, we propose a shape function $b(r)\equiv r_{0}e^{(-r/r_{0}+1)}$. For this proposal of shape function, the metric (\ref{TraversibleWormholeMetric}) takes the form
\begin{eqnarray} \label{TraversibleWormholeMetricSpecialcase}
ds^{2}&\equiv& -e^{2\Phi(r)}dt^{2}+\frac{dr^{2}}{1-r_{0}e \frac{e^{-r/r_{0}}}{r}}+r^{2}d\theta^{2}+r^{2}sin^{2}\theta d\zeta^{2}\nonumber\\
\end{eqnarray}
as general case. One can detect that, for our choice of shape function here, the second term in the denominator of the radial component of the metric has the same characteristic with the Yukawa's potential, where $r_{0}$ and $e$ are constants and are the width of the throat of the wormhole and Euler's number, respectively. In this manner, we call the wormhole described by (\ref{TraversibleWormholeMetricSpecialcase}) as the Yukawa type wormhole. One can also observe that, if $r\rightarrow\infty$ the metric (\ref{TraversibleWormholeMetricSpecialcase}) asymptotically converges to the flat, Minkowski metric for $\Phi=0$. In other words, the metric has the same geometric features with the Minkowski metric, at far distances from the physical event, the wormhole. In addition, the proposal of the Yukawa Type wormhole is important in the manner of the fact that it directly fits to the charge manifestation of the topology with its characteristic same as the characteristic of one of the most fundamental electromagnetic interaction potential. Space, which basically resembled a sheet with a handle, is nothing but that electric charge manifestation of the topology where the wormhole provides a mechanism for charge without charge \cite{NECconditions}.

Therefore, for the shape function $b(r)\equiv r_{0}e^{(-r/r_{0}+1)}$, for the $f(\phi)=\phi$ case, the condition (\ref{NECconditionOne}) becomes
\begin{equation} \label{NECconditionOneYukawa}
\rho+p_{t}-\frac{(3r_{0}+r)e^{(-r/r_{0}+1)}}{6\kappa r^{3}}\geq 0
\end{equation}
where $(3r_{0}+r)e^{(-r/r_{0}+1)}$ is definitely positive, hence
\begin{equation} \label{NECconditionOneYukawaTwo}
\rho+p_{t}\geq 0
\end{equation}
and condition (\ref{NECconditionTwo}) becomes
\begin{equation} \label{NECconditionTwoYukawa}
\rho+p_{r}\geq -\frac{(3r_{0}+r)e^{(-r/r_{0}+1)}}{3\kappa r^{3}}
\end{equation}
allowing both of the $\rho$ and the $p_{r}$ to be negative at the same time. However, this case forces $p_{t}$ to be positive due to the condition (\ref{NECconditionOneYukawaTwo}).
In other words, in order that the NEC is satisfied, both of $\rho$ and the $p_{r}$ can be negative, however $p_{t}$ is positive in that situation.
On the other hand, $p_{t}\geq p_{r}$ according to equation (\ref{PressureDifferenceSpecialCase}).
Those are actually general cases for all shape functions for which the expression $(-3b+b^{(1)}r)$ is negative.

\subsection{\label{sec:level2}{Determination of $p_{r}$ and $p_{t}$ }}

From the conservation of energy $\nabla_{\mu}T^{\mu\nu}=0$ \cite{CarollLectureNotes} for $\nu=1$ and $\Phi=0$ we get
\begin{equation} \label{RadialPressureEquation}
p^{(1)}_{r}+2r^{-1}(p_{r}-p_{t})=0
\end{equation}
equation for the pressures. However, $(p_{r}-p_{t})$ has already been introduced in equation (\ref{PressureDifferenceGeneral}). Thus, for zero tidal force case, we obtain
\begin{equation} \label{RadialPressureEquationII}
p^{(1)}_{r}=\frac{\phi}{\kappa f(\phi)} \frac{(3b-b^{(1)}r)}{r^{4}}
\end{equation}
equation for radial pressure. Hence we can determine the radial pressure as below
\begin{equation} \label{RadialPressureDet.}
p_{r}=\frac{\phi}{\kappa f(\phi)} \int^{r}_{r_{0}} \frac{3b(y)-b^{(1)}(y)y}{y^{4}}dy
\end{equation}
in zero tidal force situation. On the other hand, we can also determine the transverse pressure as
\begin{equation} \label{TransversePressureDet.}
p_{t}=p_{r}+\frac{\phi}{f(\phi)}\frac{(3b-b^{(1)}r)}{2\kappa r^{3}}
\end{equation}
in terms of the radial pressure, by means of the equation  (\ref{PressureDifferenceGeneral}), for $\Phi=0$. Furthermore, in the existance of a scalar field that corresponds to dilaton, pressures are defined as in the above equations (\ref{RadialPressureDet.}) and  (\ref{TransversePressureDet.}) however, with
$f(\phi)=\phi$ condition.

\section{\label{sec:level1}{conclusion}}
In this work, we have discussed the Morris-Thorne type wormhole, in scalar tensor theory. First, we have reached the equations of motion in the existence of a time dependent scalar field that permits a dynamic energy momentum distribution altough the metric is static, for the general case for which there is no fixed $\Phi(r)$ and/or $b(r)$ in the Scalar Tensor Model, in the presence of anisotropic distribution. Then, we have represented that the energy momentum tensor can be expressed in terms of pressure measured in only one direction. Here, we have benefited from the fact that the radial pressure and the transverse pressure are seen not to be independent of each other. In addition, their difference can be expressed in terms of the field $\phi$, a general function of the field $f(\phi)$ of the model (\ref{Action Functional}) and the redshift function $\Phi(r)$ and the shape function $b(r)$ of the wormhole. After that, we have moved to the special case $\Phi(r)=0$, zero tidal force case. In that situation, we have shown that
there is no traversable wormhole in the presence of isotropic fluid which is already a special case of the anisotropic one. Then, discussing the tree level string dilaton case benefiting from the reference \cite{ScalarFieldMatterLagrangian} we have argued the pressures and the NEC conditions in terms of the shape function of the wormhole. In the last steps of this paper, we have proposed a shape function $b(r)\equiv r_{0}e^{(-r/r_{0}+1)}$ for which the $b(r)/r$ term in the metric tensor acts as the Yukawa's potential. We noticed that, $p_{t}\geq p_{r}$ for that type. We have also discussed the NEC for this type of wormhole individually. We have concluded by this discussion that the energy density and the radial pressure are allowed to be negative at the same time in order that the NEC is satisfied. Actually, ``those situations hold for all shape functions for which $(-3b+b^{(1)}r)$ is negative", although the requirements for the NEC is given as $\rho+p_{t}\geq 0$ and $\rho+p_{r}\geq 0$ in \cite{NECconditions}. Of course, there is no contradiction between the conditions in \cite{NECconditions} and the conditions derived in this paper, but the condition on $\rho+p_{r}$ is seen to be extended for particular cases of our work.
However, the situation in which both of the $\rho$ and the  $p_{r}$ are negative is not allowed also in our study, if the signatures of the radial and the transverse pressures are the same. In addition, we can assert that (\ref{NECconditionOneYukawa}) and (\ref{NECconditionTwoYukawa}) represents the NEC conditions for the Yukawa Type wormhole, for $\Phi(r)=0$ in the presence of a dilaton like field. Furthermore, we have determined the radial and the transverse pressures in terms of the shape function $b$, scalar field $\phi$  and the general function $f(\phi)$ of the field, for zero tidal force situation in the formalism of our paper.

\section{\label{sec:level1}{acknowledgement}}

We are very grateful to Altu\u{g} \"{O}Özpineci for useful discussions.


\begin{thebibliography}{99}


\bibitem{GTRinconsistent}
K. Bhattacharya and B. R. Majhi,  Phys. Rev. D {\bf95},  064026 (2017),\\
arXiv: 1702.07166v2

\bibitem{CanOnlyBeExplainedinGTR}
V. Faraoni and S. D. Belknap-Keet,  Phys. Rev. D {\bf96},  044040 (2017),\\
arXiv: 1705.05749v1


\bibitem{ExtensionOfGTRScalarTensorTheory}
T. E. Raptis and F. O. Minotti,     Class. Quantum Grav. {\bf30},  235004 (2013),\\
arXiv: 1305.7141v1


\bibitem{ScalarTensorTheoryLowEnergyLimit}
D. Anderson, N. Yunes and E. Barausse, Phys. Rev. D  {\bf94}, 104064  (2016),\\
arXiv: 1607.08888v1

\bibitem{ScalarTensorTheoryLowEnergyString}
M. Horbatsch, H. O. Silva, D. Gerosa, P. Pani, E. Berti, L. Gualtieri and U. Sperhake, Class. Quantum Grav. {\bf32}, 20, 204001 (2015),\\
arXiv: 1505.07462v3

\bibitem{BDreference}
S. M. M. Rasouli, A. H. Ziaie, S. Jalalzadeh and P. V. Moniz, Annals of Physics 375, 154 (2016),\\
                        arXiv:1608.05958v2

\bibitem{ModelIndependentPredictionofStringTheories}
L. J. Garay and J. Garcia-Bellido, Nucl.Phys. B400, 416-434, (1993),\\
                        arXiv:gr-qc/9209015v2

\bibitem{ScalarFieldMatterLagrangian}
O. Minazzoli and A. Hees, Phys. Rev. D {\bf90}, 023017 (2014),\\
                        arXiv:1404.4266v2

\bibitem{EllisBronnikovWormhole}
X. Y. Chew, B. Kleihaus and J. Kunz, Phys. Rev. D {\bf94}, 104031 (2016),\\
                        arXiv:1608.05253v1


\bibitem{EinsteinRosenBridge}
T. Ohgami and N. Sakai, Phys. Rev. D {\bf94}, 064071 (2016),\\
                        arXiv:1704.07093v1

\bibitem{Morris-Thorne}
M. S. Morris and K. S. Thorne, Am. J. Phys. {\bf56}, 395 (1988)

\bibitem{AnisotropicDistributionTensor}
N. M. Garcia and F. S. N. Lobo, Phys. Rev. D {\bf82}, 104018 (2010),\\
arXiv:1007.3040v2

\bibitem{PastStudiesFirst}
K.A. Bronnikov and S.V. Grinyok, Grav. Cosmol. {\bf10}, 237 (2004),\\
arXiv:gr-qc/0411063

\bibitem{PastStudiesSecond}
K.A. Bronnikov, M.V. Skvortsova and A.A. Starobinsky, Grav. Cosmol. {\bf16}, 216-222, (2010),\\
arXiv:1005.3262v3

\bibitem{PastStudiesThird}
R. Shaikh and S. Kar, Phys. Rev. D {\bf94},  024011  (2016),\\
arXiv:1604.02857v3

\bibitem{PastStudiesFourth}
R. Shaikh and S. Kar, Phys. Rev. D {\bf96}, 044037 (2017),\\
arXiv:1705.11008v2

\bibitem{PastStudiesFifth}
E. F. Eiroa and G. F. Aguirre, Phys. Rev. D {\bf94}, 044016 (2016),\\
arXiv:1605.07089v3

\bibitem{PastStudiesSixth}
Z. Yousaf, M. Ilyas and M. Zaeem-ul-Haq Bhatti, Eur. Phys. J. Plus {\bf132}, 268 (2017)\\

\bibitem{PastStudiesSeventh}
Zhan-Feng Mai and H. L\"{u}, Phys. Rev. D {\bf95}, 124024 (2017),\\
arXiv:1704.05919v2

\bibitem{CarollLectureNotes}
S. M. Caroll, Lecture Notes on General Relativity, (1997),\\
arXiv:gr-qc/9712019

\bibitem{IsotropicFluid}
M. Zubair, S. Waheed, and Y. Ahmad,  Eur. Phys. J. C {\bf76}, 444 (2016),\\
arXiv:1607.05998

\bibitem{NEC}
E. Battista, E. Di Grezia, M. Manfredonia and G. Miele, Eur. Phys. J. Plus, {\bf132}, 537  (2017),\\
arXiv:1707.01508


\bibitem{NECconditions}
M. R. Mehdizadeh and A. H. Ziaie, Phys. Rev. D {\bf95},  no.6, 064049 (2017),\\
arXiv:1704.06923v1






\end{thebibliography}
\end{document}